# Empirical Analysis of the Photoelectrochemical Impedance Response of Hematite Photoanodes for Water Photo-Oxidation

Dino Klotz[1,2], Daniel A. Grave[1], Hen Dotan[1], Avner Rothschild[1]

[1] *Department of Materials Science and Engineering, Technion - Israel Institute of Technology, Haifa 32000, Israel*

[2] *International Institute for Carbon-Neutral Energy Research, Kyushu University, 744 Motooka, Nishi-ku Fukuoka 819-0395, Japan*

**Abstract**
Photoelectrochemical impedance spectroscopy (PEIS) is a useful tool for the characterization of photoelectrodes for solar water splitting. However, the analysis of PEIS spectra often involves *a priori* assumptions that might bias the results. This work puts forward an empirical method that analyzes the distribution of relaxation times (DRT), obtained directly from the measured PEIS spectra of a model hematite photoanode. By following how the DRT evolves as a function of control parameters such as the applied potential and composition of the electrolyte solution, we obtain unbiased insights into the underlying mechanisms that shape the photocurrent. In a subsequent step, we fit the data to a process-oriented equivalent circuit model (ECM) whose makeup is derived from the DRT analysis in the first step. This yields consistent quantitative trends of the dominant polarization processes observed. Our observations reveal a common step for the photo-oxidation reactions of water and $H_2O_2$ in alkaline solution.

**Table of Contents Graphic**

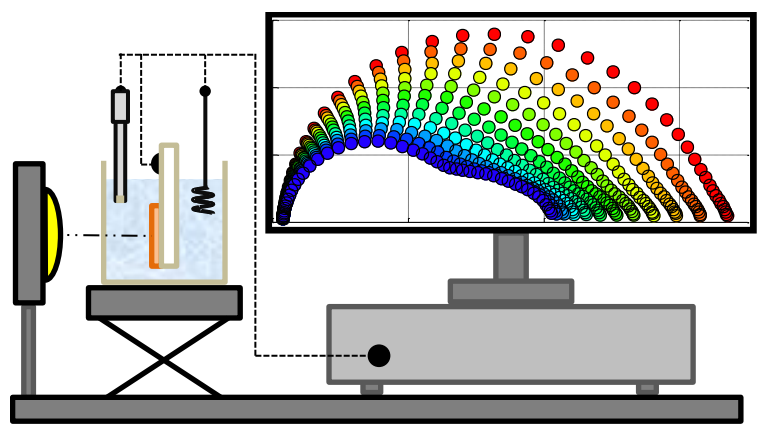

Photoelectrochemical impedance spectroscopy (PEIS) is a widely used method[1] to characterize photoelectrode materials for solar water splitting, such as hematite ($\alpha$-$Fe_2O_3$),[2–8] $BiVO_4$[9,10] and $TiO_2$.[11,12] Under constant illumination and potential bias, the dynamic photocurrent response to a small signal potential perturbation with different frequencies is measured.[13] Hematite is particularly suitable for systematic analysis, as it is one of the most stable materials for solar water splitting.[14] There is a well-established equivalent circuit model (ECM) for hematite photoanodes that was presented by Bisquert, Hamann and co-workers.[2] Its theoretical basis is provided in Ref. [15]. Recently, the model was expanded to describe the results of photoelectrochemical immittance triplets (PIT),[13] consisting of PEIS and intensity modulated photocurrent/voltage spectroscopy (IMPS/IMVS).[16] This model is able to simulate the typical shapes of PIT spectra for different operating conditions on the basis of only six parameters and is in line with the empirical approach for the analysis of PIT presented by our group.[13]

Another important challenge is to consistently explain the evolution of the ECM parameters and correlate them to the underlying polarization processes. Complementary techniques such as pump-probe optical measurements,[17–21] infrared spectroscopy[22] or microwave conductivity measurements[23] can help to clarify the nature of these processes. This work puts forward an empirical method that analyzes the distribution of relaxation times (DRT), obtained from PEIS measurements on an epitaxially grown hematite thin film photoanode.[24] The analysis identifies distinct features in PEIS spectra measured in alkaline solutions with and without sacrificial hole scavenger ($H_2O_2$),[25] and quantifies their contribution to the photocurrent by a simple process-oriented ECM. By following the evolution of these features as a function of the applied potential, we find that the photo-oxidation reactions of water and $H_2O_2$ share a common step that becomes rate limiting at high anodic potentials where the photocurrent-potential ($JU$) voltammograms with and without $H_2O_2$ coincide. These observations add new insight into the role that $H_2O_2$ plays as a sacrificial hole scavenger that is often added to the alkaline solution in order to study photoanode properties such as charge carrier separation and charge transfer efficiencies.[23]

PEIS spectra are usually displayed in a so-called Nyquist plot, where the negative imaginary part is plotted against the real part as shown in Figure 1a. Two pieces of information can be extracted directly from this plot: The Ohmic resistance corresponds to the high frequency intersect with the real axis, $R_0$, and the DC resistance corresponds to the low frequency intersect with the real axis, $R_{DC}$. From these two values the so-called polarization resistance can be calculated: $R_{pol} = R_{DC} - R_0$. In most cases, $R_0$ is typically much smaller than $R_{pol}$ and it does not depend on the potential bias or light intensity. Therefore, it is usually attributed to the resistance of the current collector. Figure 1b illustrates the relation between a PEIS spectrum and a $JU$ voltammogram. The derivative of the photocurrent with respect to potential, i.e., the slope of the $JU$ curve, is equal to $1/R_{DC}$ at any given potential bias. The smaller $R_{DC}$ the larger the slope of the $JU$ curve. Therefore, any process that increases $R_{DC}$ reduces the photocurrent $J$. Hence, the sum of all the polarization processes, $R_{pol}$, plus $R_0$ determines the shape of the $JU$ curve.

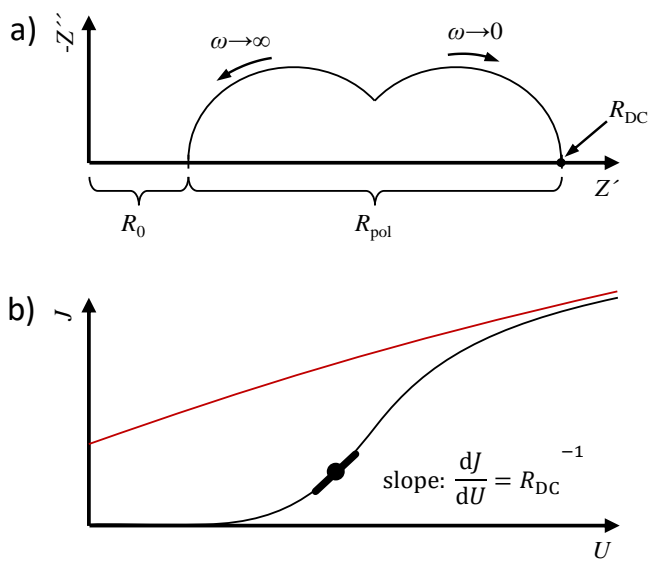

**Figure 1. (a) Schematic Nyquist plot showing two semicircles and the characteristic resistances that can be deduced directly from the diagram: $R_0$, $R_{DC}$ and $R_{pol}$ (= $R_{DC} – R_0$); (b) Schematic photocurrent-potential**

(*JU*) voltammograms for water photo-oxidation (black curve) and the corresponding hole current without surface recombination (red curve).[13,25] The black dot marks a specific operating point on the voltammogram and the black bar shows the slope at this operating point.

The characteristic features in PEIS spectra are semicircles visible between $R_0$ and $R_{DC}$. Figure 1a shows a schematic PEIS spectrum with two separate semicircles related to two distinct polarization processes with vastly different time constants, $\tau_i$, or characteristic frequencies. If the difference between the time constants of different polarization processes is smaller than one order of magnitude, the respective semicircles overlap and it becomes difficult to resolve them. Therefore, the first step in impedance analysis is deconvoluting the characteristic features in the impedance spectra. This is typically done by fitting the spectra to a pre-defined ECM, achieving two goals in one step: deconvoluting partially-overlapping features in the PEIS spectra and extracting quantitative values of the lumped elements in the ECM that represent the underlying polarization processes. The salient challenge in this approach is constructing the appropriate ECM that represents the system correctly, without subtracting or adding information to the empirical information in the measured data. This is not an easy task as the ECM relies on *a priori* assumptions that restrict the solutions to a pre-defined number of polarization processes that may or may not be appropriately represented in all measured data.

Here, we demonstrate that the distribution of relaxation times (DRT),[13,26–28] calculated directly from the PEIS spectra, can help to identify the polarization processes with minimal *a priori* assumptions. In this context, the terms "relaxation times" and "time constants" describe the same physical quantity; the different names have historic reasons, as the DRT was first derived for dielectric relaxation processes.[28] Our analysis focuses on a model hematite photoanode for water photo-oxidation, serving as a case study to demonstrate the application of the DRT analysis method and its merits in deconvoluting different polarization processes in photoelectrochemical systems. Details about the physical meaning of the DRT can be found elsewhere.[28] The DRT function, $g(\tau)$, is calculated by equation (1),

$$Z(\omega) = R_0 + \int_0^\infty \frac{g(\tau)}{1 + j\omega\tau} d\tau \quad (1)$$

using Tikhonov regularization in order to avoid false artifacts (see Supplementary Information for details).[29–31] The DRT function, $g(\tau)$, is then plotted against the frequency $f = (2\pi\tau)^{-1}$ to construct DRT spectra as shown below (see Figure 3). The DRT spectra display a more refined pattern as compared to Nyquist or Bode plots of the PEIS spectra (see Figure S1), making it easier to resolve different polarization processes. The central frequency of a peak in the DRT corresponds to the characteristic frequency of the related process, and the area under the peak is proportional to the polarization resistance of the process.[26] An ideal polarization process is represented by a Dirac pulse in the DRT.[32] However, in reality polarization processes are often non-ideal due to local variations of material properties (heterogeneity), spatio-temporal fluctuations in the reaction dynamics and other reasons.[33,34] That is why the DRT spectra usually exhibit peaks with a certain width. More details about the DRT calculation can be found in the Supplementary Information. Refs. [13,30,32,35,36] provide examples where the DRT analysis method was successfully applied to study electrochemical power devices such as fuel cells and batteries.

To demonstrate the application of the DRT analysis method to study photoelectrochemical systems, we selected a model system photoanode comprising a hetero-epitaxial hematite thin film (~30 nm) deposited on a (110) sapphire substrate with a Nb-doped $SnO_2$ (NTO) transparent electrode. Details of the fabrication process, the photoelectrochemical characteristics and the test cell can be found elsewhere.[24] 64 PEIS spectra were measured in alkaline solution (1M NaOH) without sacrificial reagents, and 9 spectra were measured with a sacrificial hole scavenger (1M NaOH + 0.5M $H_2O_2$).[25] Figure 2 shows the static current densities plotted against the applied potential obtained in PEIS measurements with $H_2O_2$ (red symbols) and without $H_2O_2$ (black symbols). The measurements were carried out under front white LED (4300K) illumination with a constant light intensity of 80 mW/cm² at the surface of the photoanode.

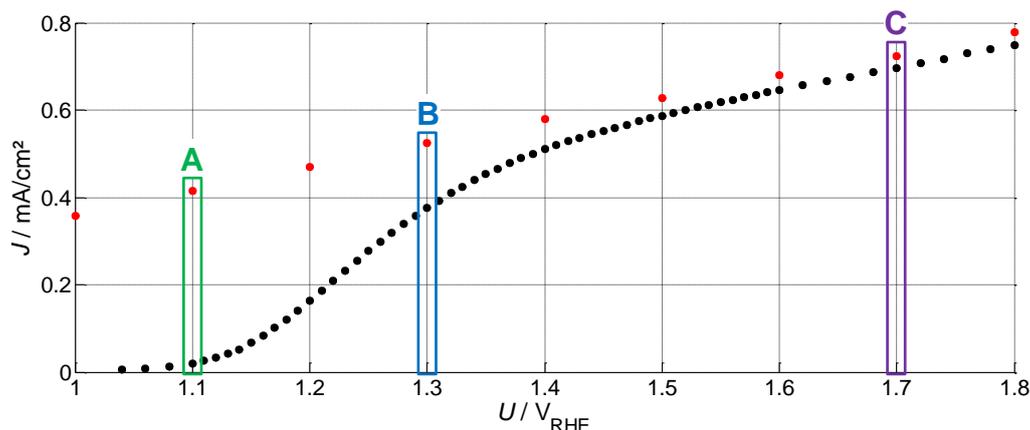

**Figure 2.** Static measurement points at which the PEIS spectra were measured in 1M NaOH (black dots) and 1M NaOH + 0.5M $H_2O_2$ (red dots) aqueous solutions under front white LED (4300 K) illumination of 80 mW/cm². A, B and C indicate the operation points for the PEIS spectra that are shown in Figure 3.

PEIS spectra were measured at the operation points depicted as dots in Figure 2. The spectra were measured using a Zahner CIMPS system and the potential was adjusted via an Hg/HgO reference electrode in 1M NaOH solution. The upper panels in Figure 3 show three exemplary PEIS spectra recorded with (red) and without $H_2O_2$ (black) at around the onset potential (1.1 $V_{RHE}$, panel A), in the plateau region (1.7 $V_{RHE}$, panel C) and in between these regions (1.3 $V_{RHE}$, panel B). The first observation is that the red spectra (with $H_2O_2$) are very similar for all these operating points. This is also demonstrated in the corresponding DRT spectra shown in the lower panels in Figure 3. The red DRT spectra, obtained from the PEIS spectra measured with $H_2O_2$, all show a similar peak frequency around 20 to 30 Hz.

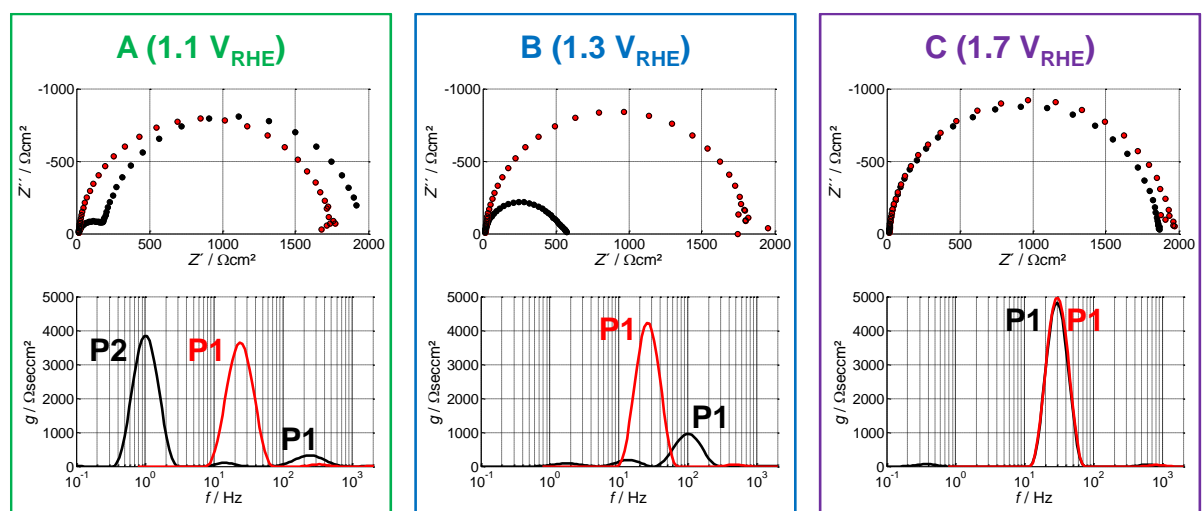

**Figure 3.** Upper panels: PEIS spectra measured at potentials of 1.1, 1.3 and 1.7 $V_{RHE}$ (A, B, and C, respectively) without (black) and with $H_2O_2$ (red). Lower panels: the corresponding DRT spectra with the same color code.

Next, we observe that the black and red PEIS spectra in panel C are quite similar. This is also demonstrated in the corresponding DRT spectra shown in the lower panel that nearly coincide, indicating almost identical characteristics of a single dominant polarization process. We conclude from this observation that there is a common polarization process, P1, in this regime that limits the photocurrent J both with and without $H_2O_2$. Also, the characteristics of the P1 peaks in panel C are almost identical and therefore, we can conclude that P1 is the same with and without $H_2O_2$ at this potential. For the PEIS spectra in panels A and B, it is not easy to draw a direct conclusion. In panel A, it seems that there is one small additional semicircle at high frequencies without $H_2O_2$ and the large semicircle looks similar for both spectra. However, the DRT spectra tell a different story. The characteristic frequencies differ by more than an order of magnitude, which is a clear indication that P1 (red) and P2 (black) originate from different polarization processes. In addition to P2, the black DRT spectrum in panel A exhibits two smaller peaks in the frequency range between 10 and 300 Hz. The larger of the two is called P1 and the smaller one is discarded as it cannot be excluded to be an artifact of the DRT calculation (see Supplementary Information for details). The pattern is consistent with the black DRT spectrum in panel B that also shows three small peaks covering a similar frequency range as in panel A.

In order to shed light on the evolution of the underlying polarization mechanisms as a function of the applied potential, we extracted the DRT spectra from all the measured (64 + 9) PEIS spectra. The results are presented in Figure 4 as color maps, mapping $g(\tau)$ as a function of frequency and applied potential. The $g(\tau)$ intensity is color coded according to the third root scale bar on the right of the figure. The data were cut below 0.3% of the highest $g(\tau)$ value in order to dispose of minor artifacts originating from the calculation of $g(\tau)$. The DRT spectra are represented by vertical lines in the $g(\tau)$ map at the respective potentials, where the y-axis shows the frequency and the color indicates the magnitude of the DRT function, $g(\tau)$. As the polarization resistance is proportional to the integrated area of $g(\tau)$ (see equation (1)), a vertical blue line in the $g(\tau)$ map indicates a negligibly small polarization resistance that goes along with a large slope of the corresponding photocurrent voltammogram at the respective potential (see Figure 1b). Points along a vertical line that tend more towards the warm (red) colors add up to a higher polarization resistance that reduces the slope in the corresponding photocurrent voltammogram at the respective potential. Hence, the $g(\tau)$ map in Figure 4a shows that the photocurrent voltammograms obtained without $H_2O_2$ are flat at potentials below 1.1 $V_{RHE}$ (i.e., left of the vertical line marked as A), and their slope becomes large at potential between 1.1 and 1.3 $V_{RHE}$ (i.e., between the vertical lines marked as A and B, respectively), and eventually the slope decreases at potentials above 1.3 $V_{RHE}$ (i.e., right of the vertical line marked as B). The $g(\tau)$ map in Figure 4b shows that the photocurrent voltammogram obtained with $H_2O_2$ displays a rather monotonic slope with a slightly convex shape over the entire potential range (1 to 1.8 $V_{RHE}$), which is represented by a similar pattern for all potentials with a slight color change towards yellow/red in the center of the colored feature marked as P1.

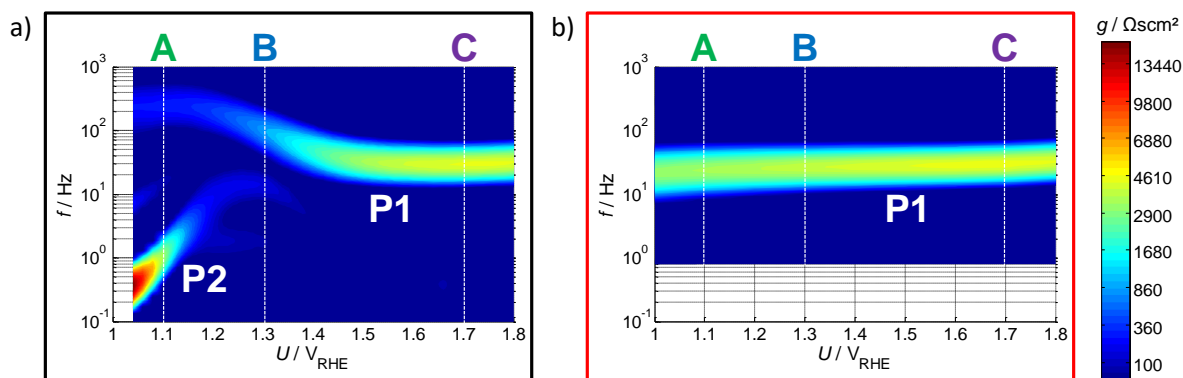

**Figure 4.** DRT maps constructed from the collection of PEIS spectra measured at the operation points shown in Figure 2. The x-axis shows the potentials at which the corresponding PEIS spectra were measured, the y-axis shows the frequency and the color indicates the magnitude of the DRT function, $g(\tau)$, at the respective potential and frequency, displayed on a third root scale. (a) Corresponds to measurements without $H_2O_2$ and (b) to measurements with $H_2O_2$.

The DRT maps in Figure 4 serve to identify the main polarization processes, their characteristic frequency, and their potential dependence trends. Figure 4(b) indicates that in operation with $H_2O_2$, there is just one dominant process, P1, that does not change significantly, neither in magnitude nor in characteristic frequency, with the applied potential. At high potentials, above ~1.5 $V_{RHE}$, there is hardly any difference in the DRT maps with and without $H_2O_2$. As already mentioned above, this indicates that at high potentials there is a common polarization process (P1) that controls the photocurrent with and without $H_2O_2$. At lower potentials (< 1.5 $V_{RHE}$), there is at least one more polarization process, P2, in the PEIS measurements without $H_2O_2$, with characteristic frequencies much lower than those of P1. There are a couple of additional processes that can be observed in Figure 4(a) but their magnitude is negligible and therefore they are discarded. At low potential (< 1.1 $V_{RHE}$), P2 clearly dominates the DRT map in Figure 4(a). In a previous study of a similar hematite photoanode, the low-frequency feature marked as P2 has been shown by empirical analysis of PIT spectra to be related to surface recombination.[13] The high-frequency feature marked as P1 dominates the PEIS spectra with $H_2O_2$ at all potentials, as well as the spectra without $H_2O_2$ at high potentials. Thus, the DRT maps in Figure 4 reveal two distinct polarization processes. One of them (P1) is common to measurements with and without $H_2O_2$ and slowly increases with increasing potentials. The other process (P2) appears only in measurements without $H_2O_2$ and unlike P1, it decreases with increasing potentials. The deconvolution of such unambiguous trends is not trivial for the studied case, especially because the characteristic frequencies almost converge when the potential approaches $U$ = 1.3 $V_{RHE}$ (B), which makes them difficult to discern with solely an ECM fitting procedure. This case study demonstrates how the DRT analysis can be applied to facilitate this task, helping to relate the features of PEIS spectra and *JU* voltammograms in a very elegant way – without *a priori* assumptions that constrain the results to a pre-defined pattern.

In order to extract quantitative information that can be linked to physical parameters such as reaction rate and pseudo capacitance, we fit the PEIS spectra to a process-oriented ECM in a subsequent step after the empirical identification of the distinct features by the DRT analysis in the first step. Herein the number of RQ elements (R – resistor, Q – constant phase element) is commensurate with the number of dominant polarization processes observed in the DRT maps in Figure 4, plus a series resistance. The obtained magnitudes and time constants can be easily cross-checked with the DRT maps in Figure 4 for consistency. For the fitting procedure, we use an algorithm that calculates the DRT of the model in each optimization step and compares it to the DRT that was extracted from the respective PEIS measurement.[36] The process-oriented ECMs with and without $H_2O_2$ are depicted in Figure 5a. The residual plots for all 64 + 9 fits can be found in the Supplementary Information, where Kramers-Kronig residuals and the residuals of the DRT calculations are presented.

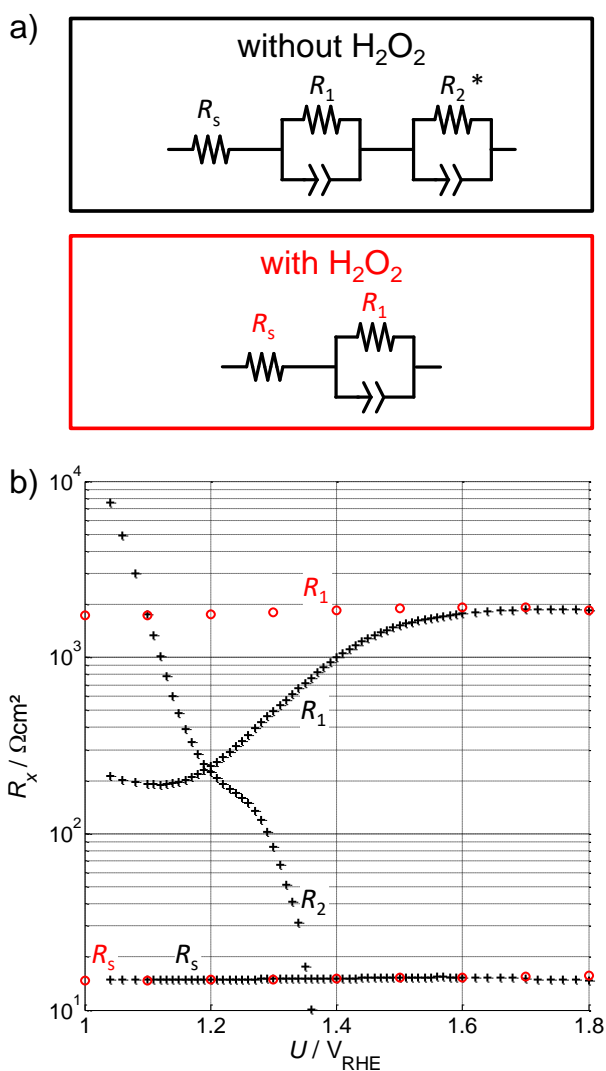

**Figure 5.** (a) Process-oriented ECMs used for fitting the PEIS spectra measured with and without $H_2O_2$ (red and black panels, respectively). $R_2^*$ was applied only at low potentials ($U < 1.4$ $V_{RHE}$). (b) Results of fitting all 64 + 9 PEIS spectra to the ECMs in (a). Red and black data points correspond to measurements with and without $H_2O_2$, respectively. The time constants obtained for the measurements without $H_2O_2$ are shown in Figure S3 in the Supplementary Information.

The trends observed in the fitted resistances in Figure 5 follow the trends in the DRT maps in Figure 4, and enable a quantitative analysis of the underlying processes. The following observations are noted:

- The series resistance $R_s$ is independent of the potential and does not change by adding $H_2O_2$ to the alkaline solution.
- $R_1$ is observed in measurements taken both with and without $H_2O_2$. In the presence of $H_2O_2$ it has a nearly constant value (1500-2000 $\Omega cm^2$), independent of the applied potential. Without $H_2O_2$ it has a low value at low potentials (~200 $\Omega cm^2$ below 1.2 $V_{RHE}$), but the value increases with increasing potentials between 1.2 and 1.6 $V_{RHE}$ and at high potentials (> 1.6 $V_{RHE}$) it reaches the same values obtained with $H_2O_2$.
- $R_2$ is observed only at low potentials without $H_2O_2$. It decreases with increasing potentials and becomes negligible for potentials above 1.4 $V_{RHE}$.

These observations can be rationalized by considering the mechanism of water photo-oxidation reported by Zandi et al.[22]. We postulate that the water and $H_2O_2$ photo-oxidation reactions share a common reaction step, the formation of double-bonded oxygen intermediate species in the first hole transfer step of the water photo-oxidation reaction, Fe-OH + OH$^-$ + $h^+$ → Fe=O + $H_2O$. If the formation of this intermediate species is slower than the successive reaction steps, it becomes the rate-limiting step for the photo-oxidation reactions both with and without $H_2O_2$. This explains the similar characteristics observed at high potentials both with and without $H_2O_2$ (compare Figures 4(b) and (a), respectively), suggesting that the formation of the double-bonded oxygen species, represented by the polarization process P1 in our DRT analysis, is the rate-limiting step at high potentials. The surface recombination reaction, Fe=O + $e^-$ + $H_2O$ → Fe-OH + $OH^-$, is only relevant without $H_2O_2$, which acts as a hole scavenger that reacts with the holes faster than their recombination with electrons at the surface.[37] Without $H_2O_2$, the Fe=O formation reaction is counteracted by the surface recombination reaction, which is represented by the polarization process P2 in the DRT analysis. The surface recombination reaction is dominant at low potentials, and decreases with increasing potentials (perhaps due to band bending and electron depletion at the surface, as suggested elsewhere[7]). Consequently, $R_2$ decreases with increasing potentials. In the presence of $H_2O_2$, the subsequent elementary steps that follow the formation of double-bonded oxygen (Fe=O) intermediate species are fast, and the surface recombination process is suppressed.

An interesting result of the analysis that warrants further explanation is the drop of $R_1$ at low potentials in measurements without $H_2O_2$. First we note that the large difference in $R_1$ in measurements with and without $H_2O_2$ indicate that the related process P1 is definitely not a pure bulk process since it changes significantly with the change in the electrolyte composition. The evolution of $R_1$ as a function of the applied potential, in measurements without $H_2O_2$, suggests that process P1, which is related to the formation of Fe=O intermediates, becomes limited at high potentials (and high photocurrents) by the availability of reaction sites at the surface. In this context it is noteworthy that Durrant and coworkers reported a change in the rate law beyond a critical potential for different photoanodes studied by photo-induced absorption measurements.[19] This observation is commensurate with saturation of the surface intermediates at high photocurrents, which is in line with our conclusion as for the limited availably of reaction sites at high potentials.

As a bottom line, the empirical DRT analysis presented here suggests a common reaction step, which is relevant for the water and $H_2O_2$ photo-oxidation reactions. This case study demonstrates the suitability of the DRT method to analyze PEIS spectra of complex photoelectrochemical reactions such as water photo-oxidation. The DRT method is very useful for empirical analysis of PEIS spectra that helps identifying distinct features in the measured spectra with minimal *a priori* assumptions. By following the evolution of these features as a function of control parameters such as potential and electrolyte composition, it might be possible to assign them to physical processes. However, the assignment requires independent verification by complementary *operando* analytical methods.[38] It is also noteworthy that only those processes with characteristic time constants that are within the frequency range of the PEIS measurements can be analyzed, whereas faster or slower processes cannot be detected by PEIS.

In summary, the photoelectrochemical impedance of photoelectrodes carries important information about the polarization processes that shape the photocurrent. However, the analysis that links this information to a mechanistic model with specific physical parameters is not a straightforward task. In many cases the analysis is biased by a pre-determined model that constrains the results to a certain pattern. In order to avoid this constrain, we propose an empirical approach to analyze PEIS measurements that consists of two consecutive steps: (i) the first step extracts the distribution of relaxation times (DRT) from a series of PEIS spectra measured at different potentials both with and without sacrificial reagents such as the $H_2O_2$ hole scavenger, followed by (ii) fitting the distinct features extracted from the DRT spectra in the first step to a process-oriented equivalent circuit model (ECM) in the second step. Applying this analysis for a case study of a model hematite photoanode in alkaline solution, we have identified two dominant polarization processes. The first process is surface recombination which is dominant at low potentials and is suppressed by adding hole scavenger ($H_2O_2$) to the solution. The second process is suspected to be the formation of double-bonded oxygen intermediates (Fe=O), which dominates the water photo-oxidation reaction at high potentials (without $H_2O_2$) and the $H_2O_2$ photo-oxidation reaction at all potentials. This work demonstrates the merits of the DRT analysis method in identifying and quantifying distinct polarization process, an important step towards understanding the reaction mechanism.

**Supporting Information Available:** Detailed information about the calculation of the DRT, examples showing the benefit of the DRT, time constants obtained for the fits shown in Figure 5b and residuals (Kramers-Kronig, DRT calculation and ECM fit).

**Author Information**
Corresponding Authors
*E-mail: avner@mt.technion.ac.il (A.R.).
*E-mail: dino.klotz@i2cner.kyushu-u.ac.jp (D.K.).

**Acknowledgement**
This research has received funding from the European Research Council under the European Union's Seventh Framework programme (FP/200702013) / ERC (grant agreement n. 617516). D.A.G. acknowledges support by Marie-Sklodowska-Curie Individual Fellowship No. 659491. The PEIS spectra were measured using central facilities at the Technion's Hydrogen Technologies Research Laboratory (HTRL), supported by the Adelis Foundation, the Nancy & Stephen Grand Technion Energy Program (GTEP) and by the Solar Fuels I-CORE program of the Planning and Budgeting Committee and the Israel Science Foundation (Grant n. 152/11).